# AI-based Clinical Assessment of Optic Nerve Head Robustness Superseding Biomechanical Testing

**Fabian A. Braeu**[1,2,3], **Thanadet Chuangsuwanich**[1,3], **Tin A. Tun**[4,5], **Alexandre H. Thiery**[6], **Tin Aung**[3,4,5], **George Barbastathis**[2,7], **and Michaël J.A. Girard**[1,5,8]

1. Ophthalmic Engineering & Innovation Laboratory (OEIL), Singapore Eye Research Institute, Singapore National Eye Centre, Singapore
2. Singapore-MIT Alliance for Research and Technology, Singapore
3. Yong Loo Lin School of Medicine, National University of Singapore, Singapore
4. Singapore Eye Research Institute, Singapore National Eye Centre, Singapore
5. Duke-NUS Graduate Medical School, Singapore
6. Department of Statistics and Applied Probability, National University of Singapore, Singapore
7. Department of Mechanical Engineering, Massachusetts Institute of Technology, Cambridge, Massachusetts 02139, USA
8. Institute for Molecular and Clinical Ophthalmology, Basel, Switzerland

**Corresponding Author:**  Michaël J.A. Girard
Ophthalmic Engineering & Innovation Laboratory (OEIL)
Singapore Eye Research Institute (SERI)
The Academia, 20 College Road
Discovery Tower Level 6,
Singapore 169856
mgirard@ophthalmic.engineering
https://www.ophthalmic.engineering

# Abstract

**Purpose:** To use artificial intelligence (AI) to: **(1)** exploit biomechanical knowledge of the optic nerve head (ONH) from a relatively large population; **(2)** assess ONH robustness from a single optical coherence tomography (OCT) scan of the ONH; **(3)** identify what critical three-dimensional (3D) structural features make a given ONH robust.

**Design:** Retrospective cross-sectional study.

**Methods:** 316 subjects had their ONHs imaged with OCT before and after acute intraocular pressure (IOP) elevation through ophthalmo-dynamometry. IOP-induced lamina-cribrosa deformations were then mapped in 3D and used to classify ONHs. Those with LC deformations superior to 4% were considered fragile, while those with deformations inferior to 4% robust. Learning from these data, we compared three AI algorithms to predict ONH robustness strictly from a baseline (undeformed) OCT volume: **(1)** a random forest classifier; **(2)** an autoencoder; and **(3)** a dynamic graph CNN (DGCNN). The latter algorithm also allowed us to identify what critical 3D structural features make a given ONH robust.

**Results:** All 3 methods were able to predict ONH robustness from 3D structural information alone and without the need to perform biomechanical testing. The DGCNN (area under the receiver operating curve [AUC]: 0.76 $\pm$ 0.08) outperformed the autoencoder (AUC: 0.70 $\pm$ 0.07) and the random forest classifier (AUC: 0.69 $\pm$ 0.05). Interestingly, to assess ONH robustness, the DGCNN mainly used information from the scleral canal and the LC insertion sites.

**Conclusions:** We propose an AI-driven approach that can assess the robustness of a given ONH solely from a single OCT scan of the ONH, and without the need to perform biomechanical testing. Longitudinal studies should establish whether ONH robustness could help us identify fast visual field loss progressors.

# Introduction

The optic nerve head (ONH) is the main site of damage in glaucoma and is exposed to a harsh biomechanical environment [1]. Specifically, it is consistently exposed to 3 major loads: the intraocular pressure (IOP; pushing the ONH posteriorly) [2], the cerebrospinal fluid pressure (CSFP; pushing the ONH anteriorly) [3], and during eye movements, by the traction from the optic nerve generating considerable shear across all ONH tissues [4]. Our group, and others, have hypothesized that excessive and cumulative deformations of the ONH from any of these loads (and their combinations) could result in glaucomatous axonal damage, either through a direct or an indirect pathway [5-7]. In essence, if one could develop a test to assess the robustness of an individual's ONH, and its sensitivity to the aforementioned loads, this may potentially help us predict who would be at risk of developing glaucoma, and for glaucoma patients, who is likely to exhibit rapid visual field loss progression. This would also motivate research to render the ONH 'indestructible' as a preventive treatment for glaucoma.

In the past, several biomechanical tests (or stress tests) have been proposed to assess the in vivo biomechanics of the human ONH. Such tests typically introduced a mechanical perturbation (e.g. a change in IOP), while the ONH was continuously imaged in three-dimensions using optical coherence tomography (OCT), or ultrasounds, to assess its mechanical response. In those studies, IOP was either modulated using **(1)** ophthalmo-dynamometry [8-11], **(2)** tight-fitting swimming goggles [12], **(3)** the water-drinking test [13], **(4)** the dark room test [14], **(5)** trabeculectomy [15, 16], or **(6)** simply using the ocular pulse as a natural load [17, 18]. Others have used the traction of the optic nerve as a mechanical perturbation [19-21]. In those scenarios, assessment of ONH robustness (or tissue stiffness) can be performed using the inverse finite element method or the virtual fields methods [22, 23].

However, none of these biomechanical tests come without drawbacks. They have yet to be optimized for their ease-of-use in a clinical setting while limiting patient discomfort. In this study, we argue that it should be possible to assess ONH robustness from a single OCT scan of the ONH, without performing a biomechanical test on the patient. The accuracy of such classifications can be enhanced by artificial intelligence (AI) if it were to learn from a large population on which biomechanical tests have already been performed.

In this study, we aimed to develop AI algorithms to: **(1)** exploit biomechanical knowledge of the ONH from a relatively large population; **(2)** assess ONH robustness from a single OCT scan of the ONH; **(3)** identify what three-dimensional (3D) structural features of the ONH make a given ONH fragile or robust. Our work may have implications for improving the diagnosis and prognosis of glaucoma and other optic neuropathies.

## Methods

**Patient Recruitment**

A total of 4,118 subjects from two cohorts were recruited at the Singapore National Eye Center (SNEC, Singapore) and retrospectively included in this study: **(1)** 336 subjects of Chinese ethnicity that underwent biomechanical testing of their ONH and OCT imaging; and **(2)** 3,782 subjects (7,531 scans) of mixed ethnicity (62% Indian and 38% Chinese) with OCT imaging only. All subjects gave written informed consent. The study adhered to the tenets of the Declaration of Helsinki and was approved by the institutional review board of the respective institutions. A summary of the patient population is shown in **Table 1**.

|  | AGE | SEX (%MALE) | NON-ROBUST ONH | ROBUST ONH | TOTAL NR. OF SCANS |
|---|---|---|---|---|---|
| **COHORT 1** | 66.1±6.7 | 55% | 162 | 174 | 336 |
| **COHORT 2** | 59.6±10.1 | 60% | - | - | 7531 |

**Table 1.** Summary of patient populations.

## Biomechanical Testing and Optical Coherence Tomography Imaging

Subjects that underwent biomechanical testing (cohort 1) had their ONH imaged with 3D spectral domain optical coherence tomography (OCT) before and after acute IOP elevation of approximately 19 mmHg from baseline through ophthalmo-dynanometry – a method to raise IOP via scleral indentation through the eyelid [4, 8, 10]. A detailed description of the procedure can be found in [9]. Additionally, subjects from cohort 2 had either one or both ONHs imaged at baseline IOP only. All subjects were seated in a dark room and if necessary, their pupil was dilated with tropicamide 1% solution. OCT scans (horizontal raster scans) covered the whole ONH region and were done with the same device (Spectralis, Heidelberg Engineering, Germany). Each OCT volume consisted of 97 B-scans (approximately 35 µm between B-scans) with 384 A-scans per B-scan (approximately 11.5 µm between A-scans) and 496 pixels per A-scan (3.87 µm between two pixels) covering a rectangular area of 15° x 10° centered on the ONH. Signal averaging, eye tracking, and the enhanced depth imaging modality of the Spectralis OCT device were used during image acquisition.

## Definition of ONH Robustness

In this study, a robust ONH is thought to be less sensitive to changes in IOP and would thus exhibit relatively small lamina cribrosa (LC) strains (i.e. deformations) due to an increase in IOP. The inverse (i.e. relatively large LC strains) would be true for a fragile ONH. This definition of ONH robustness is based on the hypothesis that excessive deformation of the LC, due to e.g. an increase in IOP, can lead to glaucomatous axonal damage either through a direct or indirect pathway [7, 24].

To extract the 3D deformation of the ONHs of cohort 1 (those that underwent biomechanical testing via an acute elevation in IOP), we used a commercial digital volume correlation (DVC) module (Amira (version: 2020.3), Thermo Fisher Scientific, USA). Details about the DVC algorithm can be found in [20]. From the extracted 3D deformation map, we then derived the volume-averaged effective strain in the LC ($E_{eff}$). The effective strain is a scalar value summarizing the complex 3D strain state [25, 26]. Based on the effective LC strain, we finally split the biomechanical data set in robust and fragile ONHs (using a threshold of $E_{eff} = 4\%$ that approximately corresponds to the median of the data set). The approach is summarized in **Figure 1**.

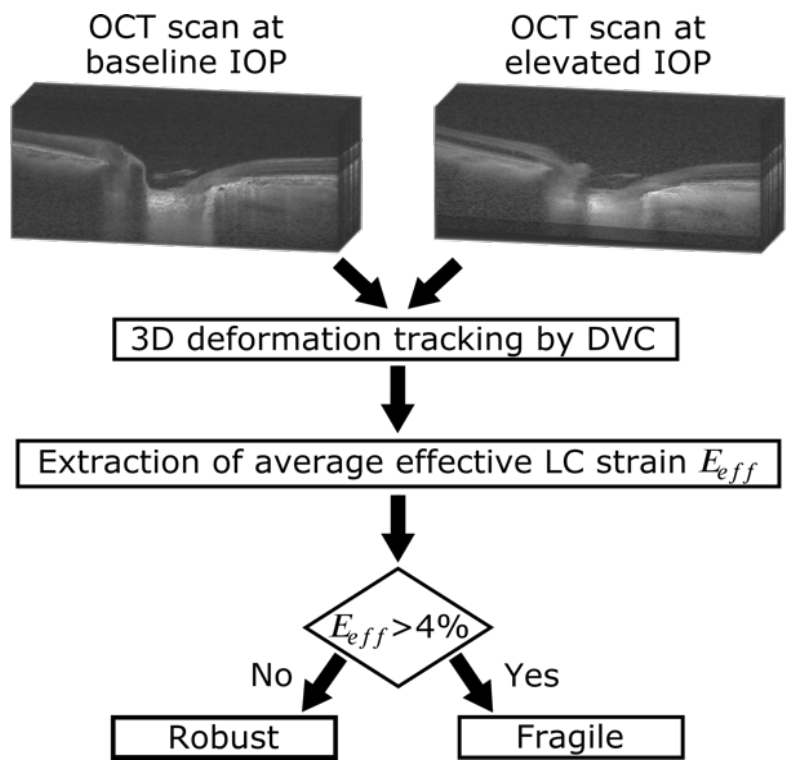

**Figure 1.** Split of biomechanical data set in robust and frail ONHs. A digital volume correlation (DVC) algorithm was applied to measure ONH deformation due to an increase in IOP. Based on the median average effective LC strain ($E_{eff} \approx 4\%$), each ONH from cohort 1 was labeled as robust ($E_{eff} < 4\%$) or frail ($E_{eff} > 4\%$).

## Algorithms to Predict ONH Robustness without Performing Biomechanical Testing

In the following sections, we present three AI algorithms (with gradually increasing complexity) that were able to learn from the already-acquired biomechanical data (from cohort 1). Such algorithms aimed to predict ONH robustness solely from a single OCT scan of the ONH without the need for any additional hardware or observations of mechanical deformations. The first algorithm was a random forest classifier – a well-established machine learning algorithm in the field of ophthalmology [27, 28]. The second was an autoencoder – a method that gained popularity in recent years [29, 30]. It was tuned herein with additional unlabeled data (cohort 2) to improve the final classification performance. The third was a dynamic graph convolutional neural network (DGCNN) [31] – a deep neural network from the group of geometric deep learning algorithms. Recently, this method was also used to identify critical 3D structural features of the ONH for glaucoma diagnosis [32].

### Algorithm 1 for AI-based Robustness Assessment – Random Forest Classifier

We trained a random forest classifier, originally proposed by Breiman in 2001 [33], to discriminate robust from fragile ONHs solely based on their structural characteristics as extracted from the baseline (undeformed) OCT volumes (cohort 1).

**Automated Segmentation.** We first segmented raw OCT volumes using the software REFLECTIVITY (Reflectivity, Abyss Processing Pte Ltd, Singapore) that was developed from advances in AI-based ONH segmentation [34, 35]. We took all baseline (undeformed) OCT volumes of cohort 1 and automatically classified the following ONH tissue groups: **(1)** the retinal nerve fiber layer (RNFL) and the prelamina tissue (PLT); **(2)** the ganglion cell inner plexiform layer (GCL+IPL); **(3)** all other retinal layers (ORL); **(4)** the retinal pigment epithelium (RPE) with Bruch's membrane (BM) and the BM opening (BMO) points; **(5)** the choroid; **(6)** the OCT-visible part of the peripapillary sclera including the scleral flange; and **(7)** the OCT-visible part of the LC.

**Automated Extraction of ONH Structural Parameters.** The following ONH structural parameters (for both neural and connective tissues) were automatically extracted from the segmented OCT volumes of cohort 1 using the software REFLECTIVITY: **(1)** the average minimum rim width in each octant defined as the minimum distance from a BMO point to a point on the inner limiting membrane (ILM); **(2)** the average RNFL thickness in each octant measured at a distance equal to 1.5 times the radius of BMO from the center of BMO; **(3)** the average ganglion cell inner plexiform layer (GCL+IPL) thickness in each octant evaluated at the same location than that used for RNFL thickness; **(4)** the prelamina depth defined as the distance from the BMO center to a point on the ILM (perpendicular to the BMO plane); **(5)** the minimum prelamina thickness; **(6)** the LC depth defined as the distance from the BMO plane to a point on the anterior LC boundary; **(7)** the LC global shape index that summarizes the shape of the anterior LC boundary into a single number [36]; and **(8)** the BMO area.

**Robustness Assessment.** All ONH structural parameters were used as inputs for the random forest classifier in order to assess whether a given ONH was robust or fragile. Optimal performance was achieved with a random forest consisting of 100 individual decision trees.

## Algorithm 2 for AI-based Robustness Assessment: Autoencoder as Classifier

We used an autoencoder (first introduced by Rumelhart et al. in 1986 [37]) in combination with a multilayer perceptron (MLP) classification network to distinguish robust from fragile ONHs solely using one segmented OCT B-scan (central slice) for each ONH.

**Automated Segmentation of Central OCT B-scans.** We automatically segmented all baseline (undeformed) OCT volumes of cohorts 1-2 with the same segmentation approach as that presented above. Subsequently, we selected the segmented OCT B-scan that was closest to the BMO center (defined as the segmented central OCT B-scan).

**Robustness Assessment.** We first trained the autoencoder (without the MLP network) in an unsupervised manner using the segmented central OCT B-scan of each OCT volume from cohort 2. To assess the performance of the autoencoder, we reported the Dice coefficient as mean ± standard deviation of all reconstructed OCT B-scans from the test set (**Figure 2a**). We then discarded the decoder and fixed the weights of the encoder. Finally, the MLP network in combination with the encoder (fixed weights) was trained in a supervised manner to predict ONH robustness from a segmented central OCT B-scan of the baseline (undeformed) OCT volume from cohort 1 (**Figure 2b**). Details about the architecture of the autoencoder and the MLP network can be found in [30]. In contrast to [30], optimal performance was achieved with a compressed representation of dimension D=64.

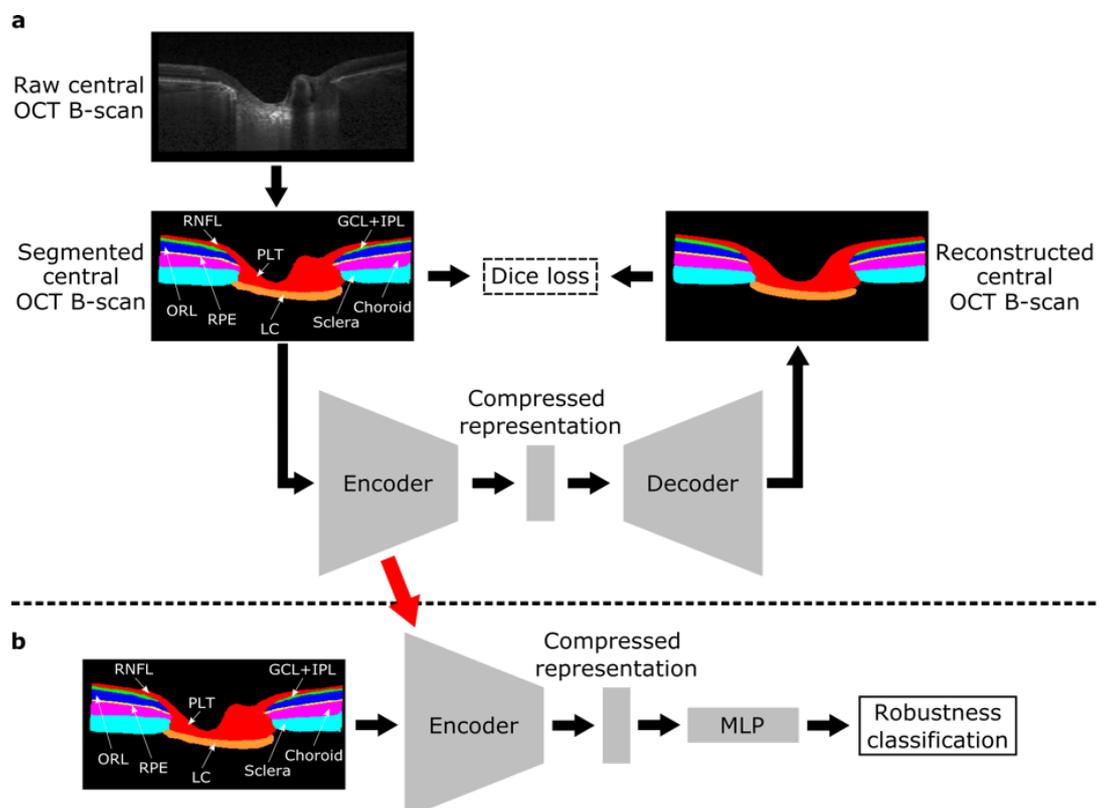

**Figure 2.** Autoencoder as a classifier. (a) The autoencoder is trained in an unsupervised manner to learn a compressed representation of a segmented central OCT B-scan. (b) The compressed representation is used as input to a MLP network to predict ONH robustness.

**Algorithm 3 for AI-based Robustness Assessment – Dynamic Graph Convolutional Neural Network**

DGCNN was used to predict whether a given ONH was considered robust or fragile. Contrary to the other two algorithms, DGCNN can learn from complex 3D shapes, as long as they are represented as 3D point clouds. Our goal was to describe a given ONH as a 3D point cloud (from the OCT scan) which then served as an input for DGCNN.

**Representing a Given ONH as a 3D Point Cloud.** From the ONH segmentations (see section above and **Figure 3a, 3b**), we identified the anterior boundaries of all aforementioned tissues, plus the posterior boundaries of the sclera and LC. About 20,000 points were randomly selected from those boundaries to represent the final 3D point cloud (**Figure 3c-d**). In addition, for each tissue, the local thickness was mapped (minimum distance between anterior and posterior boundaries) and values were assigned to points from the anterior tissue boundary. In total, each point was assigned 4 values: its coordinates [x,y,z] and a local tissue thickness (whenever applicable, or 0 otherwise). To homogenize the data, the origin of the coordinate system [x=0, y=0, z=0] was situated at the center of BMO and the BMO plane (best-fit plane to the BMO points) was aligned horizontally. For a more detailed description, the interested reader is referred to our previous publication on geometric deep learning for glaucoma diagnosis [32].

**Robustness Assessment.** DGCNN was specifically designed to process 3D point clouds (or unordered sets of points) such as those represented in **Figure 3d**. By dynamically computing local graphs for each point in the input point cloud (through edge convolutions or "EdgeConv"), the DGCNN can capture fine structural features of an object, such as small changes in curvature (**Figure 3d**). To predict robustness from a given ONH point cloud, we used the same DGCNN architecture as in [31], except that we used a max pooling layer of dimension 256 and set the number of k-nearest neighbors for all EdgeConv layers to 20.

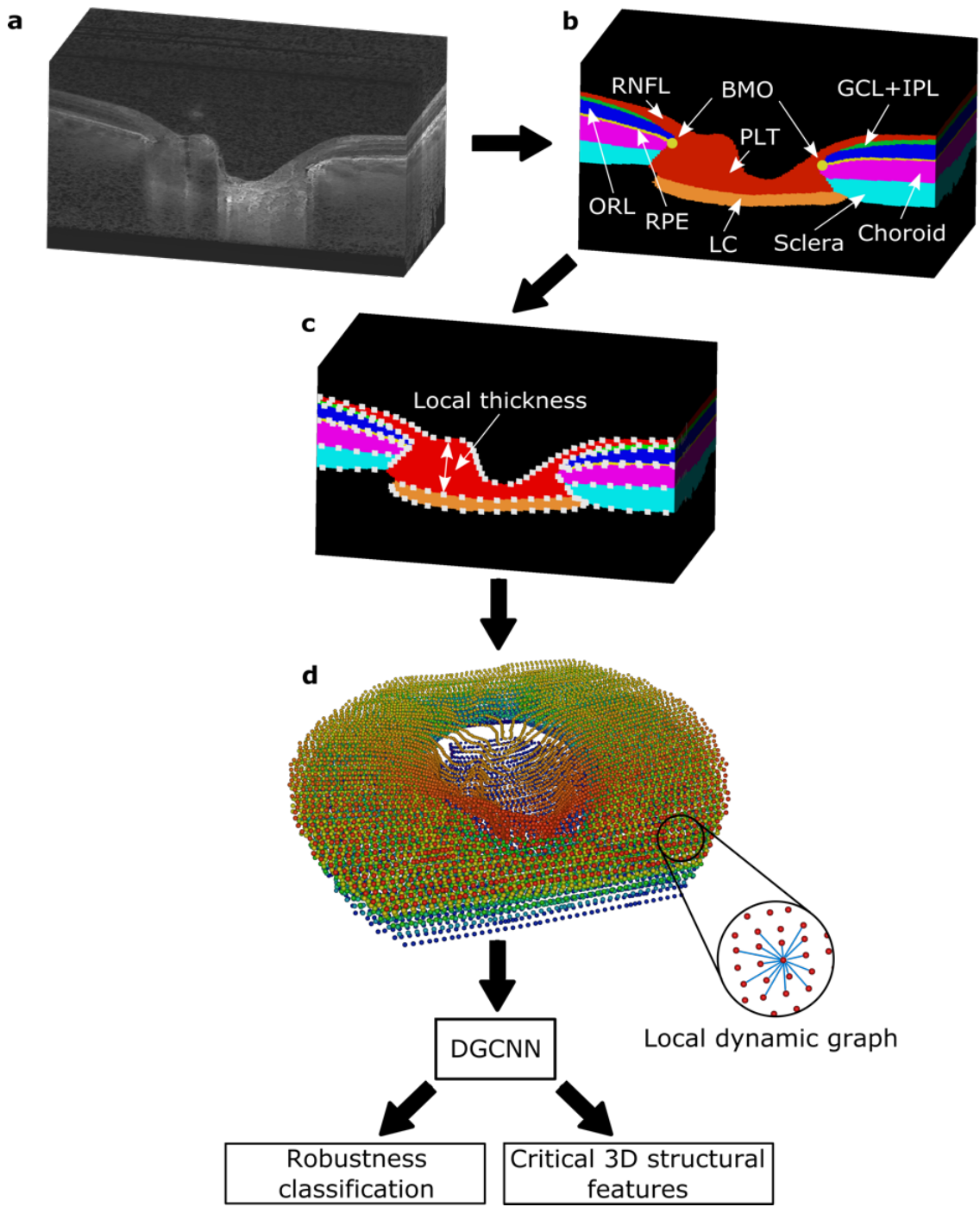

**Figure 3.** Automated extraction of 3D point clouds from raw OCT volumes representing the complex 3D structure of the ONH (a-d). Subsequently, the DGCNN was trained on these point clouds to predict ONH robustness and identify critical 3D structural features of the ONH for robustness classification.

**Identification of Critical 3D Structural Features of the ONH.** The specific architecture of the DGCNN inherently allowed us to identify regions of the ONH critical for assessing ONH robustness by extracting all points that contributed to the final classification score – the so-called critical points. We extracted the critical points for all ONHs in the best performing test set. All points were pooled and visualized as a 3D density map. A high density was obtained when a given point had many neighbours (within a 75 µm radius sphere). Since all ONHs were aligned with respect to specific landmarks (BMO plane and center), such a density map should highlight the critical ONH regions or landmarks that make a given ONH robust or fragile.

## Performance Comparison: Random Forest, Autoencoder and Dynamic Graph Convolutional Neural Network

To assess and compare the overall performance of the three classification algorithms, the biomechanical data set was split in training (70%), validation (15%), and test (15%) sets, respectively. A five-fold cross validation study was performed and we reported the receiver operating characteristic (ROC) curves and the area under the receiver operating characteristic curves (AUCs) as mean ± standard deviation. Each of the three methods used the same split of data and was trained on a Nvidia RTX A5000 GPU card until optimum performance was reached in the validation set.

To improve performance and reduce overfitting, we used data augmentation techniques such as random cropping, random rotations, random rigid translations, random sampling (i.e. randomly picking a subset of points from the input point cloud), and additive Gaussian noise where applicable.

# Results

## Assessment of ONH Robustness: Performance Comparison

All three AI-based methods were able to assess ONH robustness from a single 3D OCT scan. The DGCNN (AUC: 0.76 ± 0.08) outperformed the autoencoder (AUC: 0.70 ± 0.07) and the random forest classifier (AUC: 0.69 ± 0.05). The corresponding ROC curves are shown in **Figure 4**. Furthermore, the autoencoder was able to accurately reconstruct a segmented central OCT B-scan with a Dice coefficient of 0.91 ± 0.03.

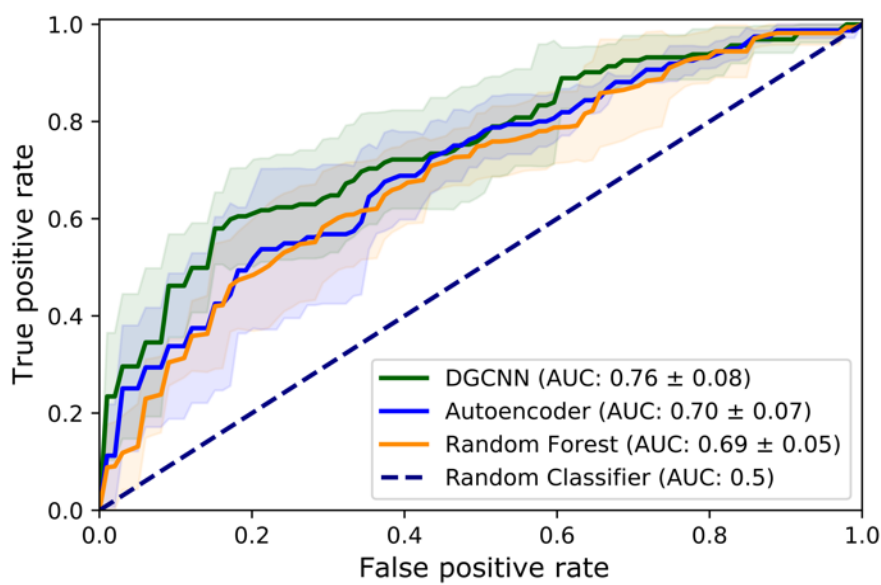

**Figure 4.** Average receiver operating characteristic curves ± standard deviation (shaded area) for the task of ONH robustness assessment with the random forest classifier (orange), the autoencoder (blue), and the DGCNN (green).

## Critical 3D Structural Features of the ONH for the Assessment of ONH Robustness

The DGCNN allowed us to directly extract critical points that were important for the assessment of ONH robustness, thus increasing the interpretability of the method. We pooled all critical points from all ONHs of the best performing test set and displayed them as a density map (**Figure 5**). The critical points formed a circle around the optic disc with points mainly located near the connective tissues (peripapillary sclera and LC) in the deeper part of the ONH. More precisely, we found that points accumulated around the scleral canal and near the LC insertion sites.

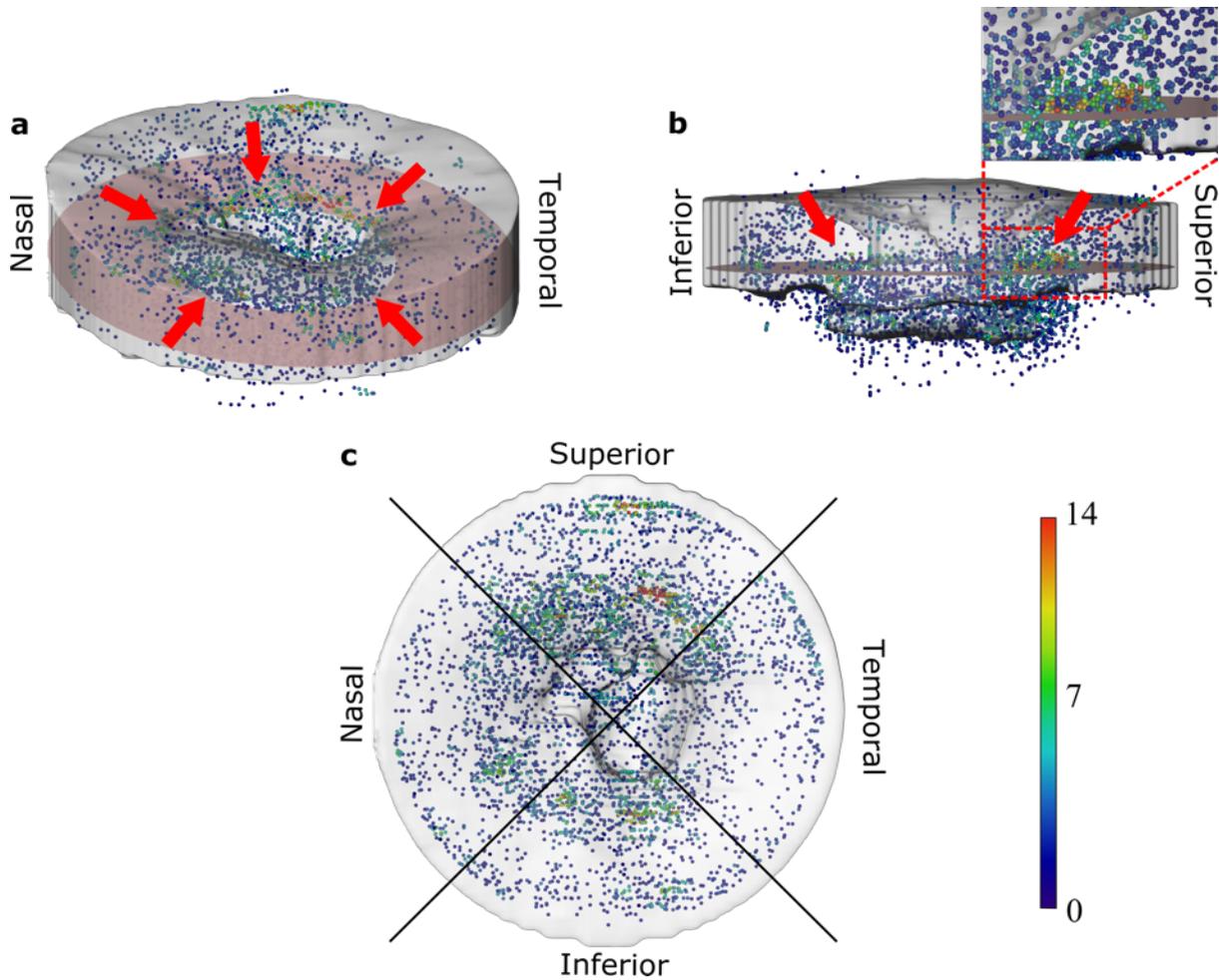

**Figure 5.** 3D (a), side (b), and top view (c) of the point cloud density plot representing the critical ONH regions for robustness assessment. The pink plane corresponds to the average anterior boundary of the sclera of all ONHs (test set) and point colours indicate the number of neighbouring points within a sphere with a radius of 0.075 mm.

## Discussion

In this study, we evaluated three AI approaches that can predict the robustness of a given ONH solely from information derived from an OCT scan. This was done without the need to perform any form of biomechanical testing on the unseen OCT volumes. Among the three AI approaches, DGCNN provided the best performance. It had the advantage of identifying what key structural landmarks (or critical points) most contributed to make a given ONH robust. Most critical points formed a ring that closely matched the location of the scleral canal boundary, while others were found near the LC insertion sites. With such a technology, our goal is to understand whether ONH robustness could help us identify glaucoma patients at risk of fast visual field loss progression.

In this study, we provided a proof-of-principle for assessing ONH robustness solely from information extracted from a single OCT scan. Specifically, it appears that ONH morphology can code for its biomechanical behavior, but most likely through an intricate relationship. Our work is a preliminary step and provides a foundation to ultimately refine such a relationship. Overall, our concept is not unfamiliar to the field of biomechanics; for instance, it has been shown that it was possible to predict the apparent elastic modulus (i.e. elastic stiffness) of trabecular bone simply from a quantitative CT scan (through assessment of the apparent density) and without performing biomechanical testing [38]. Biomechanical finite element models have also predicted that a simple change in tissue morphology could considerably alter the biomechanical environment of the ONH [39-41]. In this context, it is thus not surprising that AI might be able to pick up such relationships.

To predict robustness, we found that geometric deep learning provided the best performance. Due to the vast amount of data in a raw 3D OCT scan of the ONH, several machine learning based algorithms considered reducing the amount of input data by: **(1)** extracting characteristic structural parameters of the ONH [42]; **(2)** using two-dimensional OCT B-scans instead of the whole OCT volume [43]; or **(3)** down-sampling the raw 3D OCT scans [44]. All these pre-processing methods involve an inherent loss of fine local 3D structural features of the ONH. In contrast, the DGCNN takes a 3D point cloud as input that can represent the complex 3D structure of the ONH in an efficient way without losing fine local structural features, such as small changes in curvature. This could explain the boost in performance of DGCNN (AUC: 0.76 ± 0.08) as compared to the autoencoder (AUC: 0.70 ± 0.07) or random forest (AUC: 0.69 ± 0.05).

Overall, for robustness classification, we found that our AUCs remained in the acceptable range from 0.7 to 0.8 [45]. While the AUCs may appear low, it should be emphasized that it would almost be impossible for a human observer to discriminate robust from fragile ONHs based on a standard 3D OCT scan. Additionally, the performance of our AI algorithms might be negatively influenced by the way we created the ground truth. The median of the average effective LC strain ($E_{eff} = 4\%$) was used as the cut-off value to discriminate robust from fragile ONHs (binary classification). However, more than 50% of the subjects in the biomechanical test dataset (cohort 1) exhibited an average effective LC strain between 3% and 5%, close to the cut-off value of 4%. This alone could have strongly impacted the AUC values. In the near future, we could consider predicting the 'true' LC strains, instead of a binary classification, as this might give us a boost in performance.

Interestingly, to decide whether a given ONH was robust, our geometric deep learning algorithm especially used information from the scleral canal and LC insertion sites. Previous studies have reported the presence of a highly anisotropic (circumferentially aligned) ring of collagen fibers in the scleral flange surrounding the LC [46, 47], but also with the presence of through-thickness variations [48]. Computational studies suggested that this is an optimal fiber alignment protecting the LC from large deformations due to e.g. a change in IOP [49-51]. Therefore, pathological or age-related alterations of the collagen microstructure in the peripapillary sclera/scleral canal might have a major impact on LC strains – our definition of ONH robustness. Additionally, local LC defects or alterations like posterior movement of the LC insertion zones [52] and LC disinsertions [53] were observed in glaucomatous eyes. Prior to these defects, the ONH tissues of these glaucomatous eyes often underwent excessive remodeling leading to a change in the biomechanics of the ONH and its robustness [54]. Therefore, the position and the structure of the LC insertion sites might be a good indicator for ONH robustness.

In this study, we defined ONH robustness based on LC effective strains alone for several reasons. First, we used strain because it is a quantity that can be measured experimentally (as opposed to stress) in humans [3, 12, 16]. In a controlled biomechanical test environment with the application of a constant force (like in our study), strain also grossly reflects the structural stiffness of a structure as it is both affected by morphology and biomechanical properties. Second, we used effective strain for simplicity because it nicely summarizes the 3D state of local deformations into a single number [55]. Third, we focused on the LC because it is the major known site of axonal damage in glaucoma [56]. Note that we also fully acknowledge that ONH robustness may require an improved definition, and we hope to be able to refine it once we better understand how the ONH responds to IOP, optic nerve traction, and CSFP in a longitudinal context.

Whether ONH robustness could be used to predict visual field loss progression remains controversial. From a pure biomechanical point of view, it makes sense: a given ONH exhibiting more strains in its LC may be at greater risk of visual field loss progression. But this assumption somehow underpins the complex tissue remodeling changes that are known to occur with age, and the development and progression of glaucoma. With age, the sclera gets stiffer [57], which may in some configurations reduce LC strains. With the development and progression of glaucoma, the ONH exhibits a hyper-compliance phase in the early stages [58], followed by a stiffening phase in the more advanced stages [59, 60]. Taken together, these results, although only evaluated in primates and human donor eyes, suggest strong variations in ONH robustness throughout the course of the disease. This may need to be considered when making a prediction about visual field loss progression. We aim to better understand such implications by applying our technology to a longitudinal cohort.

In this study, several limitations warrant further discussion. First, the number of patients that underwent biomechanical testing (cohort 1) was rather small for training. In addition, the reported AUCs are only valid to this specific population and may not translate to others. In the future, it will be crucial to validate our approach in a much larger and heterogenous population. Second, our work was only tested with data acquired from one OCT device (Spectralis, Heidelberg Engineering, Germany). We were able to process OCT scans (all raster, but not diagonal) with varying resolutions and field of views, however, in the future, we must make our approach device-agnostic to be universally applicable. Third, the accuracy of the extracted point clouds might be sensitive to the performance of the segmentation algorithm. This, in turn, might influence our ONH robustness predictions. Sensitivity analyses would need to be performed to better address this issue. Fourth, our segmentations excluded some potentially relevant tissues, including, but not limited to: retinal layers (other than RNFL, GCL and IPL), the central retinal vessel trunk (and its branches) [39], and the border tissues of Elschnig and Jacoby [61]. A proper detection and subsequent representation of these tissues in the 3D ONH point cloud may in turn improve the assessment of ONH robustness. Finally, we labeled each ONH as robust or fragile according to a binary classification scheme. Instead, a "robustness score" ranging from 0 (fragile) to 1 (robust) might be more attractive for clinical translation, and this should be considered in our next iteration.

In conclusion, we have proposed an AI-driven approach that can assess the robustness of a given ONH solely from a single OCT scan of the ONH, and without the need to perform biomechanical testing. This could be achieved because our group has gathered in vivo ONH biomechanical data in a relatively large population, and from which AI could learn from. In addition, AI suggested that regions located near the scleral canal opening or near the LC insertion sites might be important landmarks to drive ONH robustness or its fragility. In the future, we should evaluate whether ONH robustness could help us identify fast visual field loss progressors.

## Acknowledgments

We acknowledge funding from **(1)** the donors of the National Glaucoma Research, a program of the BrightFocus Foundation, for support of this research (G2021010S [MJAG]); **(2)** SingHealth Duke-NUS Academic Medicine Research Grant (SRDUKAMR21A6 [MJAG]); **(3)** the "Retinal Analytics through Machine learning aiding Physics (RAMP)" project that is supported by the National Research Foundation, Prime Minister's Office, Singapore under its Intra-Create Thematic Grant "Intersection Of Engineering And Health" - NRF2019-THE002-0006 awarded to the Singapore MIT Alliance for Research and Technology (SMART) Centre [MJAG/AT/GB].